\documentclass[preprint,eqsecnum,showpacs,nofootinbib,aps,amsmath,amssymb,tightenlines]{revtex4}

\newcommand{\Lie}[0]{{\cal L}\, }
\newcommand{\grad}[0]{\nabla\!}
\newcommand{\R}{{\mathcal{R}}}
\def\th{{\widehat{\tau}}}
\def\rh{{\widehat{r}}}
\def\twoR{{\widetilde{\R}}}
\def\ls{{(\ell)}}
\def\ns{{(n)}}
\def\l{{\ell}}
\def\fie{\varphi}
\def\q{{\widetilde{q}}}
\def\K{{\widetilde{K}}}
\newcommand{\man}{{\mathcal{M}}}

\def\be{\begin{equation}}
\def\ee{\end{equation}}
\def\ba{\begin{eqnarray}}
\def\ea{\end{eqnarray}}

\preprint{\vbox{\baselineskip=12pt \rightline{gr-qc/0306115}
\rightline{CGPG-03-06/3} \rightline{NSF-KITP-03-45}}}

\begin{document}

\title{How Black Holes Grow%
\footnote{Text of a plenary talk given at the conference on
\emph{Mathematical Physics, General Relativity and Cosmology},
held in honor of Professor Jerzy Plebanski in Mexico city, in
September of 2002. To appear in the Proceedings. Results reported
here were obtained jointly with Badri Krishnan and summarized very
briefly in \cite{ak1}. Details of proofs and extensions of these
results will appear elsewhere.}
}
    \author{Abhay Ashtekar}
    \affiliation{Center for Gravitational Physics and Geometry\\
    Physics Department, Penn State University, University Park, PA 16802,
    USA;\\
    Kavli Institute of Theoretical Physics\\
    University of California, Santa Barbara, CA 93106-4030, USA}

\begin{abstract}

A summary of how black holes grow in full, non-linear general
relativity is presented. Specifically, a notion of \emph{dynamical
horizons} is introduced and expressions of fluxes of energy and
angular momentum carried by gravitational waves across these
horizons are obtained. Fluxes are local and the energy flux is
positive. Change in the horizon area is related to these fluxes.
The flux formulae also give rise to balance laws analogous to the
ones obtained by Bondi and Sachs at null infinity and provide
generalizations of the first and second laws of black hole
mechanics.

\end{abstract}

\pacs{04.25.Dm, 04.70.Bw}

\maketitle

\section{Introduction}
\label{s1}

Black holes are perhaps the most fascinating manifestations of the
curvature of space and time predicted by general relativity.
Properties of isolated black holes in equilibrium have been
well-understood for quite some time. However, in Nature, black
holes are rarely in equilibrium. They grow by swallowing stars and
galactic debris as well as electromagnetic and gravitational
radiation. For such dynamical black holes, the only known major
result in \emph{exact} general relativity has been a celebrated
area theorem, proved by Stephen Hawking in the early seventies: if
matter satisfies the dominant energy condition, the area of the
black hole event horizon can never decrease. This theorem has been
extremely influential because of its similarity with the second
law of thermodynamics. However, it is a `qualitative' result; it
does not provide an explicit formula for the amount by which the
area increases in any given physical situation. One might hope
that the change in area is related, in a direct manner, to the
flux of matter fields and gravitational radiation falling into the
black hole. Is this in fact the case? If so, the formula
describing this dynamical evolution of the black hole would give
us a `finite' generalization of the first law of black hole
mechanics: The standard first law, $\delta E = (\kappa/8\pi G)
\delta a + \Omega \delta J$, relates the infinitesimal change
$\delta a$ in the black hole area due to the infinitesimal influx
$\delta E$ of energy and angular momentum $\delta J$ as the black
hole makes a transition from one equilibrium state to a
\emph{nearby} one, while the exact evolution law would provide its
`integral version', relating equilibrium configurations which are
far removed from one another.

From a general, physical viewpoint, these expectations seem quite
reasonable. Why, then, had this question remained unresolved for
three decades?  The reason is that when one starts thinking of
possible strategies to carry out these generalizations, one
immediately encounters severe difficulties. To begin with, to
carry out this program one would need a precise notion of the
gravitational energy flux falling into the black hole. Now, as is
well known, in full general relativity, there is no gauge
invariant, quasi-local notion of gravitation radiation. The
standard notion refers to null infinity, where one can exploit the
weakness of curvature to introduce the notion of asymptotic
translations, define Bondi 4-momentum as the generator of these
translations, calculate energy fluxes, and prove the `balance law'
relating the change in the Bondi 4-momentum with the momentum flux
across portions of null infinity. \emph{Even this structure at
null infinity is highly non-trivial.} Indeed there was
considerable confusion about physical reality of gravitational
waves well into the late fifties because it was difficult to
disentangle coordinate effects from physical ones. Even Arthur
Eddington who, according to a popular legend, was `one of the only
three wise men' to understand general relativity from the very
early days, is said not to have believed in the reality of
gravitational waves. Apparently, he referred to them as pure
coordinate effects which `travelled at the speed of thought'!
Therefore, the introduction of the Bondi framework in the sixties
was hailed as a major breakthrough: it established, once and for
all, the physical reality of gravitational waves. Bondi famously
said: They are real; they carry energy; you can heat water with
them!

However, to address the issues discussed above pertaining to black
holes, one does not have the luxury of working in the asymptotic
region; one must face the highly curved geometry near black holes.
Since this is the strong field, highly non-linear regime of
general relativity, it seems hopeless to single out a translation
group unambiguously. What would the notion of energy even mean
without the usual simplifications at infinity? There do exist
formulas for the amount of  gravitational energy contained in a
given region. But typically they use pseudo-tensors and are
coordinate and/or frame dependent. Consequently, in strong
curvature regions, these expressions fail to be gauge invariant
whence, from a physical perspective, they are simply not
meaningful. Thus, even a broad conceptual framework or paradigm
was not available within which one could hope to establish a
formula relating the change in area to the flux of energy and
angular momentum falling into the black hole. The issues had
remained unresolved because it appeared that one has to
simultaneously develop the conceptual framework which is to
provide a natural home for the required notions in the strong
field regime of a black hole, \emph{and} manipulate field
equations with their full non-linearity to get explicit
expressions for energy and angular momentum fluxes.

At first these challenges seem formidable. However, it turns out
that an appropriate paradigm is in fact suggested by the strategy
used in numerical simulations of black hole formation and merger.
Using it, a program was initiated in collaboration with Badri
Krishnan \cite{ak1,bk} a few months before the Plebanskifest and
has been further developed by the two of us as well as by Ivan
Booth and Stephen Fairhurst since then. The final results
\emph{are} surprising because one can introduce the necessary
notions of energy and angular momentum fluxes in the strong field,
fully non-linear regime. But the methods used are all
well-established and quite conservative; there is nothing here
that was not available in the seventies.

The purpose of this contribution is to provide a bird's eye view
of the present situation and of prospects for near future. A
detailed and more comprehensive treatment will appear elsewhere.

\section{Conceptual Paradigm}
\label{s2}

\subsection{The idea}
\label{s2.1}

Our first task is to give a precise definition of the black hole
surface whose area is to increase during evolution.

Heuristically, one thinks of black holes as regions of space-time
from which no signal can escape to infinity. In mathematical
general relativity, this idea is captured in the notion of event
horizons. Standard space-times ---such as the Kerr family--- that
we physically think of as containing black holes, all have event
horizons. However, the notion of event horizons is extremely
non-local and teleological: it is the future boundary of the past
of future null infinity. Consequently, one can use event horizons
to detect the presence of a black hole only if one has access to
the full space-time metric, \emph{all the way to the infinite
future}. This extreme non-locality is dramatically illustrated  by
the fact that there may well be an event horizon developing in
your room as you read this page because a million years from now,
there may be a gravitational collapse in a nearby region in our
galaxy!

Because of these features, generally, the notion of an event
horizon is not very useful in practice. A striking example is
provided by numerical simulations of a gravitational collapse
leading to the formation of a black hole (or, of binary black
holes which merge). Here, one is interested in evolving suitable
initial data sets and one needs to know, at each time step, if a
black hole has formed and, if so, where it is. The teleological
nature of the event horizon makes it totally unsuitable to detect
black holes \emph{during} these evolutions. One needs a notion
that can sense the presence of the black hole and narrow down its
approximate location \emph{quasi-locally}, e.g., using only the
initial data that can be accessed at each instant of time.
Marginally trapped surfaces and apparent horizons provide such
notions. A marginally trapped surface $S$ is a 2-dimensional
sub-manifold of space-time $\man$, topologically $S^2$,  such that
the expansion of one null normal to it, say $\ell^a$, is
everywhere zero and that of the other null normal, $n^a$, is
negative. The second condition merely says that $n^a$ is the
inward pointing null normal. The non-trivial feature of $S$ is
that the expansion of the other null normal, $\ell^a$, is zero
rather than positive. Thus, none of the light rays emerging from
\emph{any} point on $S$ are directed towards the `asymptotic' or
`outside' region. Apparent horizons are associated with (partial)
Cauchy surfaces $M$. Given $M$, the apparent horizon $S$ is the
outermost marginally trapped surface lying in $M$. In numerical
simulations then, one keeps track of black holes by monitoring the
behavior of apparent horizons that emerge during the evolution.
Assuming cosmic censorship, once an apparent horizon develops,
there is an event horizon which lies \emph{outside} it.

Thus, in numerical relativity, one typically has a foliation of
the given region of the space-time $\man$ by partial Cauchy
surfaces $M$, each equipped with an apparent horizon. The apparent
horizons can `jump' discontinuously. For example, in the black
hole coalescence problem, there are two distinct apparent horizons
up to a certain time step and then there is a sudden jump to a
single connected apparent horizon. However, in practice, these
discrete jumps happen only at a few places during the course of
numerical simulation. Here, we will be concerned with those
evolutionary epochs during which the `stack of apparent horizons'
formed by evolution span a continuous world tube. Along these
apparent horizon world-tubes, the area increases and our aim is to
present formulas relating this increase with the influx of energy
and angular momentum.

Before proceeding to give precise definitions, however, let me
emphasize that the above considerations suggested by numerical
relativity serve \emph{only} as motivation. In particular, we will
not need a foliation of space-time $\man$ by partial Cauchy
surfaces $M$. The object of direct physical interest ---a
\emph{dynamical horizon}--- can be located using only the
space-time geometry, although in applications to numerical
relativity, it will typically arise as the world tube of apparent
horizons.

\subsection{Definition and methodology}
\label{s2.2}

Let us then begin with the precise definition.

\noindent \textbf{Definition:} A smooth, three-dimensional,
space-like sub-manifold $H$ in a space-time $\man$ is said to be a
\emph{dynamical horizon} if it can be foliated by a family of
two-spheres such that, on each leaf $S$, the expansion
$\theta_{(\ell)}$ of a null normal $\ell^a$ vanishes and the
expansion $\theta_{(n)}$ of the other null normal $n^a$ is
strictly negative.

Thus, a dynamical horizon $H$ is a 3-manifold which is foliated by
marginally trapped 2-spheres. Note first that, in contrast to
event horizons, dynamical horizons can be located quasi-locally;
knowledge of the full space-time is not required. Thus, for
example, you can rest assured that \emph{no} dynamical horizon has
developed in the room you are sitting in ever since it was built!
On the other hand, while in asymptotically flat space-times black
holes are characterized by event horizons, there is no one-to-one
correspondence between black holes and dynamical horizons. It
follows from properties of trapped surfaces that, assuming cosmic
censorship, dynamical horizons must lie inside an event horizon.
However, in the interior of an expanding event horizon, there may
be many dynamical horizons.%
\footnote{At this still preliminary stage, there is essentially no
control on how many dynamical horizons there can be inside an
\emph{evolving} event horizon, but the expectation is that there
will be several distinct ones.
\emph{Stationary} event horizons, by contrast, do not admit any
dynamical horizons because, for simplicity, we have tailored our
definition to cases where the area increases monotonically. In the
closely related notion of \emph{outer trapping horizons},
introduced by Hayward and discussed below, the area either
increases or remains constant. So, the event horizon of a
stationary space-time is an example of a trapping horizon.}
Nonetheless, the framework is likely to have powerful applications
to black hole physics because its results apply to \emph{each and
every} one of these dynamical horizons.

Apart from the requirement that $H$ be foliated by marginally
trapped surfaces, the definition contains only two conditions. The
first asks that $H$ be space-like. This property is implied by a
stronger but physically reasonable restriction that the `inward'
derivative ${\cal L}_n\,\theta_{(\ell)}$ of $\theta_{(\ell)}$ be
negative and the flux of energy across $H$ be non-zero. The second
condition is that the leaves be topologically $S^2$. This can be
replaced by the weaker condition that they be compact. One can
then show that the topology of $S$ is necessarily $S^2$ if the
flux of matter or gravitational energy across $H$ is non-zero.%
\footnote{This may seem surprising at first. As Carlo Rovelli
asked after my talk, in the problem of black hole merger,
initially one has two distinct horizons and finally there is only
one. So, the topology changes. How can one reconcile this with the
universality of topology of $H$? Note first that to address this
issue we should examine apparent ---not event--- horizons.
Secondly, and more importantly, by definition $H$ can be realized
only by \emph{continuous} segments of the world tubes of apparent
horizons. Before the merger, each continuous segment does have
topology $S^2\times R$ and after the merger the final world tube
has the same topology. In between the apparent horizon simply
jumps and there is no dynamical horizon.}
Thus, the conditions imposed in the definition appear to be
minimal. If these fluxes were to vanish identically, $H$ would
become isolated and replaced by a null, non-expanding horizon
\cite{afk}.

Dynamical horizons are closely related to Hayward's \cite{sh}
future, outward trapping horizons $H^\prime$. These $H^\prime$ are
3-manifolds, foliated by compact 2-manifolds $S$ with
$\theta_{(\ell)} =0$, $\theta_{(n)}<0$, and ${\cal L}_{n}\,
\theta_{(\ell)} <0$. Assuming that the null energy condition
holds, one can show that $H$ is either space-like or null. If it
is space-like, it is a dynamical horizon, while if it is null it
is a non-expanding horizon, studied extensively in \cite{afk}.
This notion is especially useful in the study of
`weakly-dynamical' horizons \cite{bf} which can be viewed as
perturbations of isolated horizons \cite{afk,prl,abl2,abl1}. The
main difference from the notion of dynamical horizons used in this
report is that while the definition of trapping horizons imposes a
condition on the derivative of $\theta_{(\ell)}$ \emph{off} $H$,
dynamical horizons refer only to geometric quantities which are
intrinsically defined on $H$. But this difference is not
physically significant because, in cases of interest, the
additional condition would probably be satisfied and dynamical
horizons will be future, outer trapping horizons. Rather, the
significant difference with respect to Hayward's work lies in the
way we analyze consequences of these conditions and in the results
we obtain. While Hayward's framework is based on a 2+2
decomposition, ours will be based on the ADM 3+1 decomposition. We
use different parts of Einstein's equations, our discussion
includes angular momentum, our flux formulae are new and our
generalization of black hole mechanics is different.

Let us begin by fixing notation. Let $\th^a$ be the unit time-like
normal to $H$ and  denote by $\grad$ the space-time derivative
operator. The metric and extrinsic curvature of $H$ are denoted by
$q_{ab}$ and $K_{ab}:={q_a}^c{q_b}^d\grad_c\th_d$ respectively;
$D$ is the derivative operator on $H$ compatible with $q_{ab}$ and
$\R_{ab}$ its  Ricci tensor. If $H$ admits more than one foliation
by marginally trapped surfaces, we will simply fix any one of
these and work with it. Thus, our results will apply to all such
foliations. Leaves of this foliation are called
\emph{cross-sections} of $H$. The unit space-like vector
orthogonal to $S$ and tangent to $H$ is denoted by $\rh^{\,a}$.
Quantities intrinsic to $S$ will be generally written with a
tilde. Thus, the two-metric on $S$ is $\q_{ab}$, the extrinsic
curvature of $S\subset H$ is
$\K_{ab}:=\widetilde{q}_a^{\,\,\,\,c}\widetilde{q}_b^{\,\,\,\,d}
D_c\rh_d$, the derivative operator on $(S, \q_{ab})$  is
$\widetilde{D}$ and its Ricci tensor is $\twoR_{ab}$. Finally, we
will fix the rescaling freedom in the choice of null normals via
$\l^a:=\th^{\,a}+\rh^{\,a}$ and $n^a:=\th^{\,a}-\rh^{\,a}$.

We first note an immediate consequence of the definition. Since
$\theta_\ls =0$ and $\theta_\ns <0$, it follows that
$$\K = \tilde{q}^{ab} D_a \rh_b = \frac{1}{2}\, \tilde{q}^{ab}
\nabla_a (\ell_b -n_b) >0. $$
Hence the area $a_S$ of $S$ increases monotonically along $\rh^{\,
a}$. Thus the second law of black hole mechanics holds on $H$. Our
first task is to obtain an explicit expression for the change of
area.

Our main analysis is based on the fact that, since $H$ is a
space-like surface, the Cauchy data $(q_{ab},K_{ab})$ on $H$ must
satisfy the usual scalar and vector constraints
\begin{eqnarray} H_S &:=& \R + K^2 - K^{ab}K_{ab}
          = 16\pi G T_{ab}\th^{\,a}\th^{\,b}
            \label{hamconstr}\\
    H_V^a &:=& D_b\left(K^{ab} - Kq^{ab}\right)
    = 8\pi G T^{bc}\th_{\, c}{q^a}_b
    \label{momconstr} \, .\end{eqnarray}
We will often fix two cross-sections $S_1$ and $S_2$ and focus our
attention on a portion $\Delta H \subset H$ which is bounded by
them.

\section{Energy fluxes and area balance}
\label{s3}

Let us now turn to the task of relating the change in area to the
flux of energy across $H$.

As is usual in general relativity, the notion of energy is tied to
a choice of a vector field. The definition of a dynamical horizon
provides a preferred direction field; that along $\ell^a$. To
extract a vector field, we need to fix the proportionality factor,
or the `lapse' $N$, let us first introduce the area radius $R$, a
function  which is constant on each $S$ and satisfies $a_S = 4\pi
R^2$. Since we already know that area is monotonically increasing,
$R$ is a good coordinate on $H$. Now, the 3-volume $d^3V$ on $H$
can be decomposed as $d^3V = |\partial R|^{-1} d^2V dR$.
Therefore, as we will see, our calculations will simplify if we
choose $N_R = |\partial R|$. Let us begin with this simple choice,
obtain an expression for the change in area and then generalize
the result to include a more general family of lapses.

Fix two cross sections $S_1$ and $S_2$ of $H$ and denote by
$\Delta H$ the portion of $H$ they bound. We are interested in
calculating the flux of energy associated with $\xi_{(R)}^a = N_R
\ell^a$ across $\Delta H$. Denote the flux of \emph{matter} energy
across $\Delta H$ by $\mathcal{F}^{(R)}_m$:
\be \mathcal{F}^{(R)}_m := \int_{\Delta H}
T_{ab}\th^{\,a}\xi_{(R)}^b d^3V.\ee
By taking the appropriate combination of (\ref{hamconstr}) and
(\ref{momconstr}) we obtain
\begin{equation} \label{fluxTr} \mathcal{F}^{(R)}_m=
\frac{1}{16\pi G} \int_{\Delta H}\, N_R \left\{H_S + 2\rh_a H_V^a
\right\}\, d^3V\, .
\end{equation}
Since $H$ is foliated by two-spheres, we can perform a $2+1$ split
of the various quantities on $H$.  Using the Gauss Codazzi
relation we rewrite $\R$ in terms of quantities on $S$:
\begin{equation} \label{threeR} \R = \twoR + \K^2 -\K_{ab}\K^{ab} +
2D_a\alpha^a \end{equation}
where $\alpha^a = \rh^{\,b}D_b\rh^{\,a} - \rh^{\,a}D_b\rh^{\,b}$.
Next, the fact that the expansion $\theta_\ls$ of $\l^a$ vanishes
leads to the relation
\begin{equation} \label{expansion0} K + \K = K_{ab}\rh^{\,a}\rh^{\,b} \,
.\end{equation}
Using  (\ref{threeR}) and (\ref{expansion0}) in (\ref{fluxTr}) and
simplifying, we obtain the result
\begin{equation} \label{NR} \int_{\Delta H} N_R \twoR\,d^3V = 16\pi G
\int_{\Delta H} T_{ab}\th^{\,a}\xi_{(R)}^b\,d^3V + \int_{\Delta H}
N_R\left\{ |\sigma|^2 + 2|\zeta|^2\right\}\,d^3V
\end{equation}
where $|\sigma|^2=\sigma_{ab}\sigma^{ab}$ with $\sigma_{ab}$ being
the shear of $\l^a$, and $|\zeta|^2= \zeta^a\zeta_a$ with
$\zeta^a:=\widetilde{q}^{\,ab}\rh^{\,c}\grad_c\l_b$; both
$\sigma_{ab}$ and $\zeta^a$ are tensors intrinsic to $S$. To
simplify the left side of this equation, recall that the volume
element $d^3V$ on $H$ can be written as $d^3V = N_R^{-1}dR\,d^2V$
where $d^2V$ is the area element on $S$. Using the Gauss-Bonnet
theorem, the integral of $N_R\twoR$ can then be written as
\begin{equation} \int_{\Delta H} N_R\twoR \,d^3V =
\int_{R_1}^{R_2} dR\left(
\oint_S \twoR\,d^2V\right) = 8\pi(R_2-R_1) \, .\end{equation}
(It is this manipulation that dictated our choice of $N_R$.)
Substituting this result in (\ref{NR}) we finally obtain
\be \label{balance1} \left(\frac{R_2}{2G}- \frac{R_1}{2G}\right) =
\int_{\Delta H} T_{ab}\th^{\,a}\xi_{(R)}^b\,d^3V \, +\,
\frac{1}{16\pi G} \int_{\Delta H} N_R\left\{ |\sigma|^2 +
2|\zeta|^2\right\} \,d^3V . \ee
This is the first key result we were looking for. Let us now
interpret the various terms appearing in this equation. The left
side gives us the change in the horizon `radius' caused by the
dynamical process under consideration. The first integral on the
right side of this equation is the flux $\mathcal{F}^{(R)}_m$ of
matter energy associated with the vector field $\xi_{(R)}^a$.
Since $\xi_{(R)}^a$ is null and $\th$ time-like, if $T_{ab}$
satisfies, say, the dominant energy condition, this quantity is
guaranteed to be non-negative. Since the second term is purely
geometrical and emerged as the `companion' of the matter term, it
is tempting to interpret it as the flux $\mathcal{F}^{(R)}_g$ of
$\xi_{(R)}^a$-energy in the gravitational radiation:%
\footnote{While the presence of the shear term $|\sigma|^2$ in the
flux formula (\ref{ab}) is natural from one's expectations based
on the weak field limit, the term $|\zeta|^2$ is surprising. Booth
and Fairhurst have shown that this term is of two orders higher
than the shear term in the weak field expansion. Thus, it captures
some genuinely non-linear, strong field physics which is yet to be
understood fully.}
\begin{equation} \label{ab}
 \mathcal{F}^{(R)}_g := \frac{1}{16\pi G}
 \int_{\Delta H} N_R\left\{ |\sigma|^2 + 2|\zeta|^2\right\}\,d^3V\, .
\end{equation}
Is this proposal physically viable? We will now argue that the
answer is in the affirmative in the sense that it passes all the
`text-book' tests one uses to demonstrate the viability of the
Bondi flux formula at null infinity.

First, since we did not have to introduce any structure, such as
coordinates or tetrads, which is auxiliary to the problem, the
expression is obviously `gauge invariant'. Second, the energy flux
is manifestly non-negative. Third, all fields used in it are
local; we did not have to perform, e.g., a radial integration to
define any of them. Fourth, the expression vanishes in the
spherically symmetric case: Since the only spherically symmetric
vector field and trace-free, second rank tensor field on a
2-sphere are the zero fields, if the Cauchy data $(q_{ab},K_{ab})$
and the foliation on $H$ are spherically symmetric,
$\sigma_{ab}=0$ and $\zeta^a=0$. Next, one might be concerned that
the flux may not vanish in stationary space-times. Even in the
Schwarzschild space-time, could one not construct a clever,
non-spherical dynamical horizon $H$? If one could, the area law
(\ref{ab}) would hold and then we would be led to an absurd
conclusion that there is flux of gravitational energy across this
$H$! Even if this is not possible in the Schwarzschild space-time,
could it not happen in a more general stationary space-time? If it
can, the `gravitational energy-flux' interpretation would not be
viable. Now, since $H$ is foliated by marginally trapped surfaces,
it follows from general results that it must lie inside the event
horizon. Using the fact that there is a  Killing field in that
region, one can show that there are no dynamical horizons in this
interior region, whence the concern is unfounded. Thus, the
expression on the right side of (\ref{ab}) shares with the
Bondi-Sachs energy flux at null infinity all its key properties.
We will therefore interpret it as the $\xi_{(R)}$-energy flux of
carried by gravitational waves. Recently, Booth and Fairhurst
\cite{bf} have verified that on `weakly-dynamical' horizons, the
expression reduces to the familiar one from perturbation theory.
They have also shown that this formula can be derived from a
Hamiltonian framework where $H$ is treated as the inner boundary
of the space-time region of interest. These results provide
considerable further support for our interpretation. Nonetheless,
it is important to continue to think of new criteria and make sure
that (\ref{ab}) passes these tests.

\emph{Remark}: The emergence of a precise formula for the flux of
energy across $\Delta H$ is very surprising. What would happen if
we repeat the above procedure for a general space-like surface
$\tilde{H}$? The analog of the flux term would be much more
complicated and \emph{fail to be positive definite}. This happens
even if we assume that $\tilde{H}$ is foliated by strictly
---rather than marginally--- trapped surfaces $\tilde{S}$, i.e. if we
replaced the condition $\sigma_{(\ell)} =0$ by $\sigma_{(\ell)}
<0$. Thus, there is no satisfactory candidate for the flux formula
across $\tilde{H}$. To summarize, although the calculation is
straightforward, it crucially depends on subtle cancellations
which occur \emph{precisely} because $H$ is a dynamical horizon.

To conclude this section, let us discuss the possibility of
choosing more general lapse functions. In the above calculation,
we needed a specific form of the lapse to cast $N_R d^3V$ into the
form $d^2V dR$. This suggests that we use more general functions
$r$ which are constant on each leaf $S$ of the foliation and set
$N_r = |\partial r|$. If we use a different radial coordinate
$r^\prime$, then the lapse is rescaled according to the relation
\begin{equation} N_{r^\prime} =  \frac{dr^\prime}{dr} \, N_r\, .
\end{equation}
Thus, although the lapse itself will in general be a function of
all three coordinates on $H$, the relative factor between any two
permissible lapses can be a function only of $r$. Recall that, on
an isolated horizon, physical fields are time independent and null
normals ---which play the role of $N_r \ell^a$ there--- can be
rescaled by a positive constant \cite{afk,abl2}. In the present
case, the horizon fields are `dynamical', i.e., $r$-dependent, and
the rescaling freedom is by a positive function of $r$. Thus, the
freedom in the choice of lapse is just what one would expect.

Given a lapse $N_r$, following the terminology used in the
isolated horizon framework, the resulting vector fields
$\xi^a_{(r)}:= N_r\l^a$ will be called \emph{permissible}. By
repeating the above calculation, it is easy to arrive at a
generalization of (\ref{ab}) for any permissible vector field:
\be \label{balance2} \left(\frac{r_2}{2G}- \frac{r_1}{2G}\right) =
\int_{\Delta H} T_{ab}\th^{\,a}\xi_{(r)}^b\,d^3V \, +\,
\frac{1}{16\pi G} \int_{\Delta H} N_r\left\{ |\sigma|^2 +
2|\zeta|^2\right\} \,d^3V \, , \ee
where the constants $r_1$ and $r_2$ are values the function $r$
assumes on the fixed cross-sections $S_1$ and $S_2$. This
generalization of (\ref{balance1}) will be useful in section
\ref{s5}.

\section{Angular momentum}
\label{s4}

To obtain the integral version of the full first law, we need the
notion of angular momentum and angular momentum flux. It turns out
that the angular momentum analysis is rather straightforward and
is, in fact, applicable to an arbitrary space-like hypersurface.
Fix \emph{any} vector field $\fie^a$ on $H$ which is tangential to
the cross-sections of $H$. Contract $\fie^a$ with both sides of
(\ref{momconstr}). Integrate the resulting equation over the
region $\Delta H\subset H$, perform an integration by parts and
use the identity $\Lie_\fie q_{ab} = 2D_{(a}\fie_{b)}$ to obtain
\begin{eqnarray} && \label{balanceJ} \frac{1}{8\pi G}\oint_{S_2}K_{ab}
\fie^a\rh^{\,b} \, d^2V - \frac{1}{8\pi G}
\oint_{S_1}K_{ab}\fie^a\rh^{\,b} \, d^2V  \nonumber \\
&& = \int_{\Delta H} \left( T_{ab}\th^{\,a}\fie^b + \frac{1}{16\pi
G} P^{ab}\Lie_\fie q_{ab}\right)\, d^3V\end{eqnarray}
where $P^{ab}:=K^{ab}-Kq^{ab}$. It is natural to identify the
surface integrals with the generalized angular momentum
$J^{(\fie)}$ associated with those surfaces and set:
\begin{equation} \label{jdynamic}J_S^{(\fie)} =
-\frac{1}{8\pi G} \oint_{S} K_{ab} \fie^a\rh^{\,b} \, d^2V
\end{equation}
where we have chosen the overall sign to ensure compatibility with
conventions normally used in the asymptotically flat context. The
term `generalized' emphasizes the fact that the vector field
$\fie^a$ need not be an axial Killing field even on $S$; it only
has to be tangential to our cross-sections.

The fluxes of this angular momentum due to matter fields and
gravitational waves are, respectively,
\begin{eqnarray} \mathcal{J}^{(\fie)}_\textrm{m} &=& -\int_{\Delta H}
 T_{ab}\th^{\,a}\fie^b\, d^3V \, ,\\
 \mathcal{J}^{(\fie)}_{\textrm{g}} &=& -\frac{1}{16\pi G}
 \int_{\Delta H} P^{ab}\Lie_\fie q_{ab}\, d^3V \, , \end{eqnarray}
and we get the balance equation
\begin{equation}  J_2^{(\fie)} - J_1^{(\fie)} =
 \mathcal{J}^{(\fie)}_\textrm{m}+
\mathcal{J}^{(\fie)}_{\textrm{g}}\, . \end{equation}
As expected, if $\fie^a$ is a Killing vector of the three-metric
$q_{ab}$, then the gravitational flux vanishes:
$\mathcal{J}^{(\fie)}_{\textrm{g}} = 0$.  For the discussion of
the integral version of the first law, it is convenient to
introduce the \emph{angular momentum current}
\be{j}^{\fie}:=-K_{ab}\fie^a\rh^{\,b}\ee
so that the angular momentum formula becomes
\be J_S^{(\fie)}= \frac{1}{8\pi G} \,\oint_S {j}^{\fie}\, d^2V\,
.\ee

\section{Finite version of the first law}
\label{s5}

Let us now combine the results of sections \ref{s3} and \ref{s4}
to obtain the physical process version of the first law for $H$
and a mass formula for an arbitrary cross-section of $H$.

To begin with, let us ignore angular momentum and consider the
vector field $\xi_{(R)}$ of section \ref{s3}. Denote by
$E^{\xi_{(R)}}$ the $\xi_{(R)}$-energy of cross-sections $S$ of
$H$. While we do not yet have the explicit expression for it, it
is natural to assume that, because of the influx of matter and
gravitational energy, $E^{\xi_{(R)}}$ will change by an amount
$\Delta\, E^{\xi_{(R)}} = \mathcal{F}^{(R)}_m+
\mathcal{F}^{(R)}_g$ as we move from one cross-section to another.
Then, the infinitesimal form of (\ref{balance1}), $dR/2G =
dE^{\xi_{(R)}}$, suggests that we define \emph{effective surface
gravity} $k_R$ associated with $\xi^a_{(R)}$ as $k_R:=1/2R$ so
that the infinitesimal expression is recast into the familiar form
$({k_R}/{8\pi G}) da = d E^{\xi_{(R)}}$ where $a$ is the area of a
generic cross-section. For a general choice of the radial
coordinate $r$, (\ref{balance2}) yields a generalized first law:
\begin{equation} \label{1law1} \frac{k_r}{8\pi G} \,\,da
= d E^{\xi_{(r)}}
\end{equation}
provided we define the effective surface gravity $k_r$ of
$\xi^a_{(R)}$ by
\begin{equation} k_r = \frac{dr}{dR}\,\,k_R \qquad \textrm{where}
 \qquad \xi^a_{(r)} = \frac{dr}{dR}\,\,\xi^a_{(R)} \, .\end{equation}
Note that this rescaling freedom in surface gravity is completely
analogous to the rescaling freedom which exists for Killing
horizons, or, more generally, isolated horizons \cite{afk,abl2}.
The new feature in the present case is that we have the freedom to
rescale the $\l^a$ and the surface gravity  by a function of the
radius $R$ rather than by a constant. This is just what one would
expect in a dynamical situation since $R$ plays the role of `time'
along $H$. Finally, note that the differentials appearing in
(\ref{1law1}) are \emph{actual, physical variations} along the
dynamical horizon due to an infinitesimal change in $r$ and are
not variations in phase space as in the formulations
\cite{rw,afk,abl2} of the first law based on Killing or isolated
horizons. Thus, (\ref{1law1}) is a \emph{physical version} of the
first law, whence (\ref{balance2}) is the \emph{finite version} of
the first law in the absence of rotation.

Next, let us include rotation. Pick a vector field $\fie^a$ on $H$
such that $\fie^a$ is tangent to the cross-sections of $H$, has
closed orbits and has affine length $2\pi$. (At this point,
$\fie^a$ need not be a Killing vector of $q_{ab}$.) Consider time
evolution vector fields $t^a$ which are of the form
$t^a=N_r\l^a-\Omega\fie^a$ where $N_r$ is a permissible lapse
associated with a radial function $r$ and $\Omega$ an arbitrary
function of $r$.%
\footnote{Constancy of $\Omega$ on cross-sections implies rigid
rotation, although the frequency of rotation is allowed to change
in `time'. After my talk, Alberto Garcia raised the interesting
issue of allowing differential rotations. This can be done by
letting $\Omega$ be a general function on $H$. In this case, one
can still obtain the integral version of the first law but, as one
would expect, if $\Omega$ has angular dependence, one can not
recover the familiar infinitesimal form of the first law.}
(On an isolated horizon, the analogs of these two fields are
constant.)  Evaluate the quantity $\int_{\Delta H}
T_{ab}\th^{\,a}t^b\, d^3V$ using (\ref{balanceJ}) and
(\ref{balance1}):
\begin{eqnarray}  &&\frac{r_2-r_1}{2G} + \frac{1}{8\pi G} \biggl\{
\oint_{S_2}\Omega j^\fie\,d^2V -\oint_{S_1} \Omega j^\fie\,d^2V
\biggr. - \biggl.\int_{\Omega_1}^{\Omega_2} d\Omega \oint_S
j^\fie\,d^2V \biggr\} = \nonumber\\
&& \int_{\Delta H} T_{ab}\th^{\,a}t^b \,d^3V + \frac{1}{16\pi
G}\int_{\Delta H} N_r \left(|\sigma|^2 + 2|\zeta|^2\right)\,d^3V -
\frac{1}{16\pi G} \int_{\Delta H}\Omega P^{ab}\Lie_\fie q_{ab} \,
d^3V \, .\label{integral1law}
\end{eqnarray}
This is our finite version of the familiar first laws of the
isolated horizon framework \cite{afk,abl2}. For, if we now
restrict ourselves to infinitesimal $\Delta H$, the three terms in
the curly brackets combine to give $d(\Omega J) - J d\Omega$ and
we obtain
\begin{equation} \label{genfirstlaw}  \frac{dr}{2G}+
\Omega dJ = \frac{k_r}{8\pi G}da + \Omega dJ\,= \, dE^t .
\end{equation}
This equation is just the familiar first law but now in the
setting of dynamical horizons. Since the differentials in this
equation are variations along $H$, this can be viewed as a
\emph{physical process version} of the first law. Note that for
each allowed choice of lapse $N_r$, angular velocity $\Omega(r)$
and vector field $\fie^a$ on $H$, we obtain a `permissible' time
evolution vector field $t^a=N_r\l^a-\Omega\fie^a$ and a
corresponding first law. This situation is very similar to what
happens in the isolated horizon framework \cite{afk,abl2} where we
obtain a first law for each permissible time translation on the
horizon. Again, the generalization from that time independent
situation consists of allowing the lapse and the angular velocity
to become r-dependent, i.e., `dynamical'.

For every allowed choice of $(N_r,\Omega(r),\fie^a)$, we can
integrate (\ref{genfirstlaw}) on $H$ to obtain a formula for $E^t$
on any cross section but, in general, the result may not be
expressible just in terms of geometric quantities defined locally
on that cross-section. However, in some physically interesting
cases, the expression \emph{is} local. For example, in the case of
spherical symmetry, it is natural to choose $\Omega=0$ and $R$ as
the radial coordinate in which case we obtain $E^t=R/2G$. This is
just the irreducible (or Hawking) mass of the cross-section. Even
in this simple case, (\ref{integral1law}) provides a useful
balance law, with clear-cut interpretation. Physically, perhaps
the most interesting case is the one in which $q_{ab}$ is only
axi-symmetric with $\fie^a$ as its axial Killing vector. In this
case we can naturally apply, at each cross-section $S$ of $H$, the
strategy used in the isolated horizon framework to select a
preferred $t^a$: Calculate the angular momentum $J$ defined by the
axial Killing field $\fie$, choose the radial coordinate $r$ (or
equivalently, the lapse $N_r$) such that
\begin{equation} k_r = k_o(R) :=
\frac{R^4-4G^2J^2}{2R^3\sqrt{R^4+4G^2J^2}} \end{equation}
and choose $\Omega$ such that
\begin{equation} \Omega= \Omega_o(R) := \frac{2GJ}{R\sqrt{R^4+4G^2J^2}}
 \, .\end{equation}
This functional dependence of $k_r$ on $R$ and $J$ is exactly that
of the Kerr family. With this choice of $N_r$ and $\Omega$, the
energy $E^t_S$ is given by the well known Smarr formula
\begin{equation} E^{t_o} = 2\, (\frac{k_o a}{8\pi G} + \Omega_o J) =
\frac{\sqrt{R^4 + 4G^2J^2}}{2GR} \, .\end{equation}
Thus, as a function of its angular momentum and area, each
cross-section is assigned simply that mass which it would have in
the Kerr family. This may seem like a reasonable but rather
trivial strategy. The non-triviality lies in two facts. First,
with this choice, there is still a balance equation in which the
flux of gravitational energy $\mathcal{F}^{(t_o)}_g$ is local and
positive definite (see (\ref{integral1law})). (The gravitational
angular momentum flux which, in general, has indeterminate sign
vanishes due to axi-symmetry.) Second, as mentioned in section
\ref{s3}, Booth and Fairhurst have recently shown that this
expression of the dynamical horizon energy emerges from a
systematic Hamiltonian framework on $\man$ where $H$ is treated as
an inner boundary.

Motivated by the isolated horizon framework, we will refer to this
canonical $E^{t_o}$ as the \textit{mass} associated with
cross-sections $S$ of $H$ and denote it simply by $M$. Thus, among
the infinitely many first laws (\ref{genfirstlaw}), there is a
canonical one:
\begin{equation}  dM = \frac{k_o}{8\pi G} da + \Omega_o dJ \, .
\end{equation}
Note that the mass and angular momentum depend only on geometrical
fields on each cross section and, furthermore, the dependence is
local.  Yet, thanks to the constraint part of Einstein's
equations, changes in mass over \emph{finite} regions $\Delta H$
of $H$ can be related to the expected matter fluxes and to the
flux of gravitational radiation which is local and positive.

I will conclude this section with a conceptual subtlety,
emphasized by Stephen Fairhurst in recent conferences. The first
law (\ref{integral1law}) discussed here is really a conservation
law and as such it is a finite or integral version of the first
law of black hole \emph{mechanics}. In the discussion of the first
law in its standard, infinitesimal form, one explicitly or
implicitly considers transitions between an equilibrium state and
a \emph{nearby} equilibrium state. Conceptually, this is the same
setting as in laws of equilibrium thermodynamics. In particular,
in infinitesimal processes involving black holes, the change in
surface gravity can be ignored just as the change in the
temperature is ignored in the first law $dE = TdS + W$ of
thermodynamics. During fully non-equilibrium thermodynamical
processes, by contrast, the system does not have time to come to
equilibrium and there is no canonical notion of its temperature.
Similarly, in the case of dynamical horizons, we only have a
notion of `effective' or `average' surface gravity; in striking
contrast to what happens on isolated horizons which describe
\emph{equilibrium} configurations,  $\kappa_r$ does not have the
\emph{geometrical} interpretation of surface gravity. However, if
one considers `weakly-dynamical' horizons and regards them as
perturbations of isolated horizons, there \emph{is} a geometrical
notion of surface gravity (of which $\kappa_r$ is a 2-sphere
average). In this situation, the geometrical surface gravity
appears to be a good analog of the temperature, the idea being
that system is evolving slowly so that it can reach approximate
equilibrium in spite of time dependence. In these situations, the
dynamical first law (\ref{integral1law}) can be simplified by
keeping terms only up to second order in perturbation \cite{bf}
and that expression can be regarded as the integral version of the
first law of black hole \emph{thermodynamics}.

\section{Discussion}
\label{s6}

I will conclude the report by pointing out some important open
problems whose solutions will add very significantly to our
knowledge of dynamical horizons and suggest some applications of
this framework.

i) An important open problem is to obtain a complete
characterization of the `initial data' on dynamical horizons. Can
one characterize the solutions to the constraint equations such
that $(H, q_{ab}, K_{ab})$ is a dynamical horizon? Note that this
would, in particular, provide a complete control on the geometry
of the world tube of apparent horizons that will emerge in
\emph{all possible} numerical simulations! One can further ask:
Can one isolate the freely specifiable data in a useful way? Are
these naturally related to the freely specifiable data on isolated
horizons \cite{abl1}? In the spherically symmetric case, these
issues are straightforward to address and an essentially complete
solution is known. It would be very interesting to answer these
questions in the axi-symmetric case.

ii) In the analysis of section \ref{s5}, let us drop the
restriction to $t_o^a$ and consider general permissible vector
fields $t^a$. Unlike the vector fields $\xi^a_{(r)} = N_r \ell^a$,
the vector field $t^a$ is not necessarily causal. Therefore the
matter flux $\int_{\Delta H} T_{ab}t^a \th^{\, b} d^3 V$ need not
be positive. Similarly, if $\fie^a$ is not a Killing field of
$q_{ab}$, the gravitational flux need not be positive. Therefore,
although the area $a$ always increases with $R$, $E^t$ can
decrease as $R$ increases. This is the analog of the Penrose
process in which `rotational energy' is extracted from the
dynamical horizon. Do Einstein's equations with physically
reasonable matter allow one to extract \emph{all} the rotational
energy? Once the first question in i) above is answered, one would
have an essentially complete control on such issues in \emph{fully
dynamical processes}.

iii) In a gravitational collapse or a black hole merger, one
expects the dynamical horizon in the distant future to
asymptotically approach an isolated horizon. Is this expectation
correct? If so, what can one say about the rate of approach? There
exists a useful characterization of the Kerr isolated horizon
\cite{lp}. Under what conditions is one guaranteed that the
asymptotic isolated horizon is Kerr? On an isolated horizon one
can define multipoles invariantly \cite{aaentropy} and the
definition can be carried over to each cross-section of the
dynamical horizon. Can one physically justify this generalization?
If so, what can one say about the rate of change of these
multipoles? Can one, for example, gain insight into the maximum
amount of energy that can be emitted in gravitational radiation,
from the knowledge of the horizon quadrupole and its relation to
the Kerr quadrupole? Is the quasi-normal ringing of the final
black hole coded in the rate of change of the multipoles, as was
suggested by heuristic considerations using early numerical
simulations?

iv) As the vast mathematical literature on black holes shows, the
infinitesimal version (\ref{genfirstlaw}) of the first law is
conceptually very interesting. The finite balance equation
(\ref{integral1law}) is likely to be even more directly useful in
the analysis of astrophysical situations. In particular, there
exist an infinite number of balance equations. Can they provide
useful checks on numerical simulations in the strong field regime?
Similarly, the Hamiltonian framework of Booth and Fairhurst could
be used as a point of departure for quantum mechanical treatments
beyond equilibrium situations. Can one extend the non-perturbative
quantization of \cite{abck,aaentropy} to incorporate these
dynamical situations? To naturally incorporate back reaction in
the Hawking process?

v) There has been considerable interest in the geometric analysis
community in the inequalities conjectured by Penrose, which say
that the total (ADM) mass of space-time must be greater than the
area of the apparent horizon on any Cauchy slice.  In the time
symmetric case (i.e., when the extrinsic curvature on the Cauchy
slice vanishes) this conjecture was proved last year by Huisken
and Ilmamen \cite{hi} and Bray \cite{b}. The area law of dynamical
horizons provides a nice setting to extend this analysis not only
beyond the time symmetric case but to establish a stronger version
relating the area of apparent horizons to \emph{the future limit
of the Bondi energy at null infinity}.
\bigskip

\textbf{Acknowledgements:} This report is based on joint work with
Badri Krishnan. I would also like to thank Chris Beetle, Ivan
Booth, Steve Fairhurst and Jerzy Lewandowski for stimulating
discussions, Carlo Rovelli and Alberto Garcia for raising
interesting issues during the conference, and Jim Isenberg for a
careful reading of thew manuscript. This work was supported in
part by the National Science Foundation grants PHY-0090091,
PHY99-07949, the National Science Foundation Cooperative Agreement
PHY-0114375, and the Eberly research funds of Penn State.

\end{document}